\documentclass[twocolumn,floats,prl,superscriptaddress,showpacs,aps]{revtex4}
\usepackage[dvips]{graphicx}

\begin{document}

\title{Infrared Hall effect in the electron-doped high $T_c$ cuprate Pr$_{2-x}$Ce$_{x}$CuO$_4$}
\author{A. Zimmers}
\affiliation{Center for Superconductivity Research, Department of
Physics, University of Maryland, College Park, Maryland 20742,
USA.}
\author{L. Shi,$^{a,2}$ D. C. Schmadel}
\affiliation{Department of Physics, University of Maryland,
College Park, Maryland 20742, USA.}
\author{W. M. Fisher}
\affiliation{Center for Superconductivity Research, Department of
Physics, University of Maryland, College Park, Maryland 20742,
USA.}
\author{R. L. Greene}
\affiliation{Center for Superconductivity Research, Department of
Physics, University of Maryland, College Park, Maryland 20742,
USA.}
\author{H. D. Drew}\email{hdrew@physics.umd.edu}
\affiliation{Department of Physics, University of Maryland,
College Park, Maryland 20742, USA.}
\author{M. Houseknecht}
\affiliation{Department of Physics, University at Buffalo, SUNY,
Buffalo, NY 14260, USA}
\author{}
\affiliation{Department of Physics, University at Buffalo, SUNY,
Buffalo, NY 14260, USA}
\author{G. Acbas}
\affiliation{Department of Physics, University at Buffalo, SUNY,
Buffalo, NY 14260, USA}
\author{M.-H. Kim}
\affiliation{Department of Physics, University at Buffalo, SUNY,
Buffalo, NY 14260, USA}
\author{M.-H. Yang}
\affiliation{Department of Physics, University at Buffalo, SUNY,
Buffalo, NY 14260, USA}
\author{J. Cerne}
\affiliation{Department of Physics, University at Buffalo, SUNY,
Buffalo, NY 14260, USA}
\author{J. Lin}
\affiliation{Physics Department, Columbia University, New York,
New York 10027, USA.}
\author{A. Millis}
\affiliation{Physics Department, Columbia University, New York,
New York 10027, USA.}

\date{\today}

\begin{abstract}
The electron-doped cuprate Pr$_{2-x}$Ce$_{x}$CuO$_4$ is investigated
using infrared magneto-optical measurements. The optical Hall
conductivity $\sigma_{xy}(\omega)$ shows a strong doping, frequency
and temperature dependence consistent with the presence of a
temperature and doping-dependent coherent backscattering amplitude
which doubles the electronic unit cell. The data suggest that the
coherent backscattering vanishes at a quantum critical point inside
the superconducting dome and is associated with the commensurate
antiferromagnetic order observed by other workers. Using a spectral
weight analysis we have further investigated the Fermi-liquid like
behavior of the overdoped sample. The observed Hall-conductance
spectral weight is about 10 times less than that predicted by band
theory, raising the fundamental question concerning the effect of
Mott and antiferromagnetic correlations on the Hall conductance of
strongly correlated materials.
\end{abstract}


\maketitle

Doping a Mott insulator yields a number of exotic properties due
to strong correlations in a great number of materials
\cite{Imada}. Among these, the properties of cuprates remain one
of the greatest challenges in condensed matter physics. The last
few years have seen remarkable progress on the synthesis of high
critical temperature (high T$_c)$ cuprates, followed by reliable
and high performance energy and momentum spectroscopic
measurements. It is now possible to compare quantitatively the
electron-doped side and the hole-doped side in the cuprate phase
diagram. One striking difference is the presence of
antiferromagnetic order in a much larger region on the
electron-doped side of the phase diagram. This characteristic has
been seen directly by neutron \cite{Greven1,Greven2}, $\mu$Sr
\cite{Uefuji} and indirectly by optics \cite{Zimmers} and angle
resolved photoemission (ARPES) \cite{Armitage}. However, the exact
location of the antiferromagnetic region with respect to the
superconducting region remains an open question as does the effect
of antiferromagnetic order and fluctuations on electronic
properties. Recent inelastic neutron studies \cite{Greven2} have
concluded that these two phases do not overlap by showing that the
antiferromagnetic phase is destroyed at dopings just before the
appearance of the superconducting dome. On the contrary, various
ARPES \cite{Armitage, Matsui2005, Matsui2006} and optical
\cite{Zimmers} studies have shown the presence of a large energy
pseudogap in this material. At low temperature this feature, which
may be a signature of the antiferromagnetism, persists at dopings
well inside the superconducting dome and ends at a quantum
critical point at $x_c$ = 0.165 above optimal doping. These
results are consistent with DC Hall experiments \cite{Dagan} which
have also shown strong evidence of this quantum critical point
inside the superconducting dome. At high temperatures, standard
optical measurement and ARPES data have also shown that this large
energy pseudogap disappears \cite{Zimmers, Matsui2005}.

To investigate further signatures of antiferromagnetism in the
electron-doped cuprates, we have performed infrared Hall (IR Hall)
measurements of the electron-doped Pr$_{2-x}$Ce$_{x}$CuO$_4$
(PCCO) cuprate over a wide range of dopings and temperatures. The
IR Hall response $\sigma_{xy}$ exhibits a rich structure,
including frequency and temperature dependent sign changes, which
can reveal more about the electronic structure of materials than
the longitudinal conductivity $\sigma_{xx}$. For example, IR Hall
measurements have been found to be an extremely sensitive probe to
the strong correlation effects in the hole-doped cuprates
\cite{Rigal, Grayson, Cerne1}. In this Letter we show clear
evidence at low temperatures and low dopings of a gap-like feature
well inside the superconducting dome and its closing at higher
dopings. These results support the scenario of a strong change in
the Fermi surface in the electron-doped cuprates due to
antiferromagnetic ordering and the presence of a quantum critical
point inside the superconducting dome.

\begin{figure}
\begin{center}
\includegraphics[width=5.8cm]{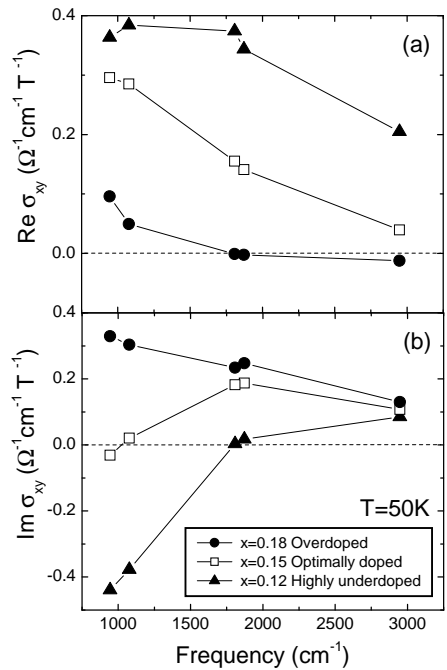}
\end{center}\caption{(a) Real and (b) imaginary part of $\sigma_{xy}$
as a function of frequency at 50K. Note the suppressed zero on the
frequency axis. The underdoped superconducting x=0.12 and
optimally doped sample x=0.15 show a clear sign change in
Im($\sigma_{xy}$). As described in the text, this feature is a
signature of antiferromagnetic order in the sample. The overdoped
superconductor x=0.18 does not show a sign change in
Im($\sigma_{xy}$).} \label{Fig1}
\end{figure}
\begin{figure}
\begin{center}
\includegraphics[width=5.2cm]{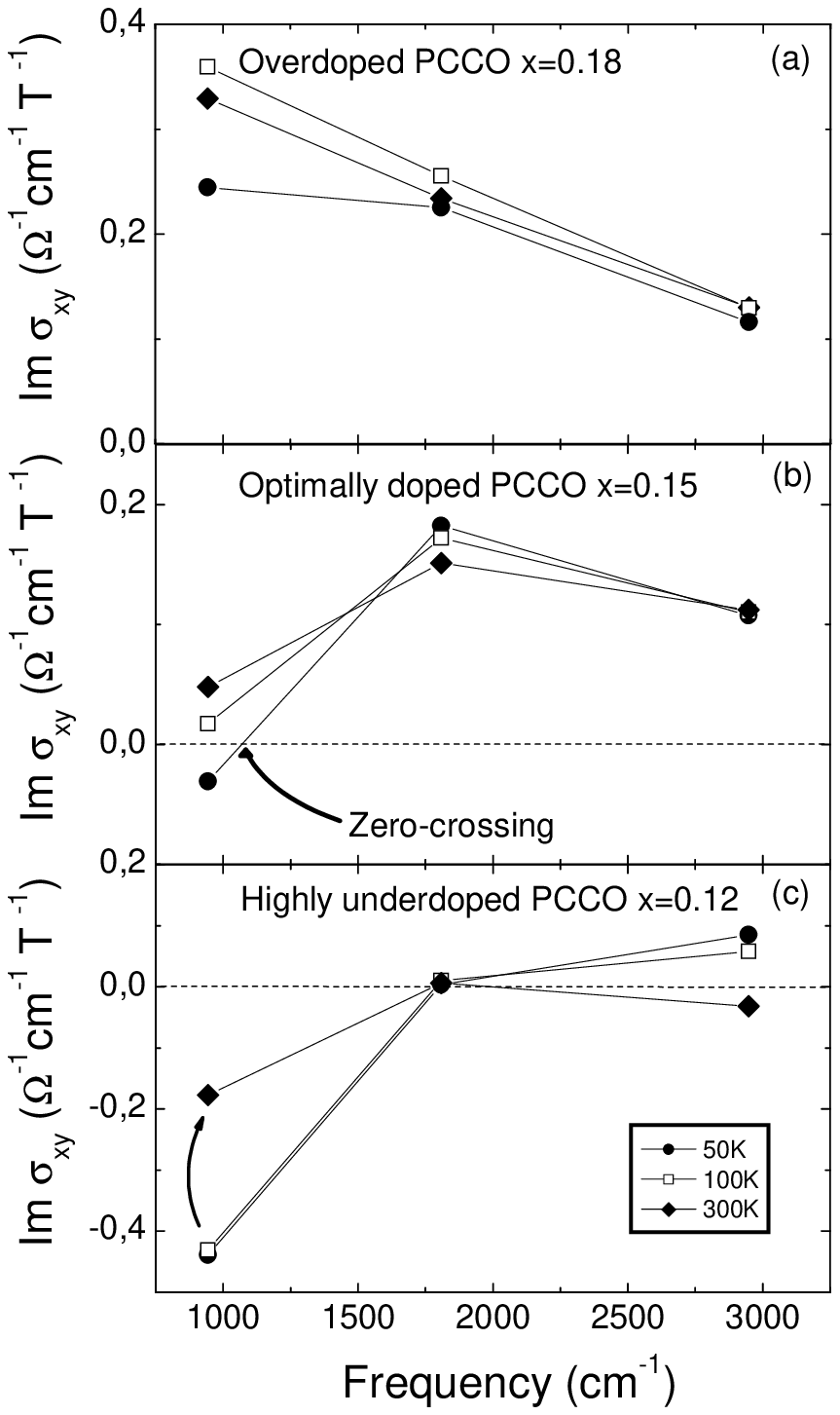}
\end{center}\caption{Temperature evolution of Im($\sigma_{xy}$) for x=0.18, 0.15 and 0.12.
As temperature is raised, the zero-crossing in this response
function clearly shifts to lower frequency for the optimally doped
sample x=0.15.} \label{Fig2}
\end{figure}

Thin films of Pr$_{2-x}$Ce$_{x}$CuO$_{4}$ of several compositions
were grown on LaSrGaO$_4$ (LSGO) substrates using pulsed laser
deposition \cite{Matsui}. In this experiment we have studied the
following samples: a highly underdoped ($x$=0.12) $T_c=2$~K
(thickness 1260~\AA), optimally doped ($x$=0.15) $T_c=19.6$~K
(thickness 1450~\AA) and overdoped ($x$=0.18) $T_c=9.3$~K
(thickness 1250~\AA). We have measured the reflectance (R),
transmittance (T) and the complex Faraday rotation angle. From
these we have extracted the longitudinal conductivity
($\sigma_{xx}$) and Hall conductivity ($\sigma_{xy}$). The
reflectance and transmittance were measured for all samples in the
400-4000~cm$^{-1}$ spectral range with a Bomem Fourier Transform
spectrometer, enabling us to determine $\sigma_{xx}$ in this
frequency range. Data were typically taken at 5 temperatures
between 30~K and 300~K using a continuous flow cryostat. The
Faraday rotation and ellipticity, represented by the complex
Faraday angle $\theta_F$, were measured using a photoelastic
polarization modulation technique \cite{Cerne3} from 30K to 300K
and from -8T to 8T. Experimental errors generated by this analysis
in the real and imaginary part of $\sigma_{xy}$ due to errors in
measurements of $\sigma_{xx}$ and $\theta_F$ are estimated to be
$\pm5\%$.

As previous studies of the longitudinal conductivity
\cite{Zimmers} have shown, the low temperature and low doping
(x=0.12) reflectance and transmittance data clearly show an extra
spectral contribution at $\approx$1500cm$^{-1}$ compared to the
standard optical response in cuprates. This spectral shape
translates into the opening of a partial gap in the optical
conductivity $\sigma_{xx}$. At low temperatures, the gap magnitude
is smaller in the intermediate doping (x=0.15) than it is at the
x=0.12 concentration.  And it is not present at all at x=0.18. At
room temperature none of the dopings presents the signature of
this spectral gap in R, T or $\sigma_{xx}$ spectral functions.

Using the measured values of $\sigma_{xx}$, the Faraday angle
$\theta_F$ can be translated into other response functions such as
the infrared Hall conductivity $\sigma_{xy}$ (the off diagonal
term in the complex conductivity tensor $\tilde{\sigma}$) or the
complex Hall coefficient $R_H$ \cite{Cerne3}. In this work we
focus on $\sigma_{xy}$ as the fundamental magneto-transport
quantity. Previous studies have validated this technique by
measuring the IR Hall response of simple metals such as gold and
copper \cite{Cerne2}. The optical properties of simple metals are
well described by a Drude model, and the IR Hall spectra for these
cases were found to be consistent with a Drude model with mass
parameter taken from band theory and measured scattering rate from
$\sigma_{xx}$.

Figure \ref{Fig1} shows our principal results: the real and
imaginary parts of $\sigma_{xy}$ as a function of frequency for
x=0.12, 0.15 and 0.18 at T=50K. We begin our discussion of these
data with reference to the Drude model with scattering rate
$\gamma$, in which $\sigma_{xy}=S_0/(\gamma-i\omega)^2$, with
$S_0$ determined by band structure. In this model
Im($\sigma_{xy}$) has the same sign at all frequencies, while
Re($\sigma_{xy}$) has one sign change at a frequency set by
$\gamma$. The data for x=0.18 are qualitatively consistent with
the Drude model: Im($\sigma_{xy}$) has the hole-like signature
predicted by the band structure at this doping, does not change
sign over the experimentally accessible frequency range, and at
low frequencies, Re($\sigma_{xy}$) has the hole-like sign found in
the DC measurement \cite{Dagan}, along with one sign change (at
$\approx$1750cm$^{-1}$) in the experimental frequency range. One
qualitative inconsistency with the Drude model is that, in the
model the maximum of Im($\sigma_{xy}$) occurs only slightly above
the frequency at which Re($\sigma_{xy}$) crosses zero. Extended
Drude models with frequency dependent scattering rates also
exhibit this feature.  However in the data the maximum of
Im($\sigma_{xy}$) obviously lies well below the zero-crossing of
Re($\sigma_{xy}$). We believe that this difference arises because
of electron-like contributions to $\sigma_{xy}$ below the
frequency range of our measurements due to magnetic fluctuation
effects.  From Kramers-Kronig relations these low frequency
contributions would increase Re($\sigma_{xy}$) while doing little
to Im($\sigma_{xy}$) in our frequency range and therefore push the
zero-crossing of Re($\sigma_{xy}$) to higher frequencies. Evidence
for these effects are also seen in the DC Hall coefficient in
similarly doped samples as multiple zero-crossings of $R_H$ with
temperature.

The data we have obtained for the lower doping x=0.15 and x=0.12
differ qualitatively from the predictions of the Drude model. In
the less doped samples Im($\sigma_{xy}$) exhibits zero-crossings
at frequencies which increases as doping decreases. While
Re($\sigma_{xy}$) is concomitantly larger (in the experimental
frequency range), DC Hall measurements indicate an electron-like
sign at both x=0.15 and x=0.12 which implies a sign change must
occur below our lowest measured frequency.

Figure \ref{Fig2} shows the temperature variation of
Im($\sigma_{xy}$) for all three dopings. As expected the overdoped
sample x=0.18 does not present a zero-crossing in this response
function at any temperature. For the optimally doped sample x=0.15
the zero-crossing clearly shifts to lower frequencies (see arrow in
panel (b)) below the measured frequency range as the temperature is
raised. This feature is not observed for x=0.12, however the low
frequency response is moving to smaller values as temperature is
raised (see arrow in panel (c)). These variations could indicate the
closing of the high energy pseudogap as seen in ARPES
\cite{Matsui2005}, $\sigma_{xx}$ \cite{Zimmers} and DC Hall
measurements \cite{Dagan}. The DC Hall and IR Hall measurements seem
however to be more sensitive than ARPES and standard optical
measurements when probing the temperature evolution of this gap
since they still clearly show the presence of the high energy
pseudogap at room temperature in underdoped samples.
\begin{figure}
\begin{center}
\includegraphics[width=9cm]{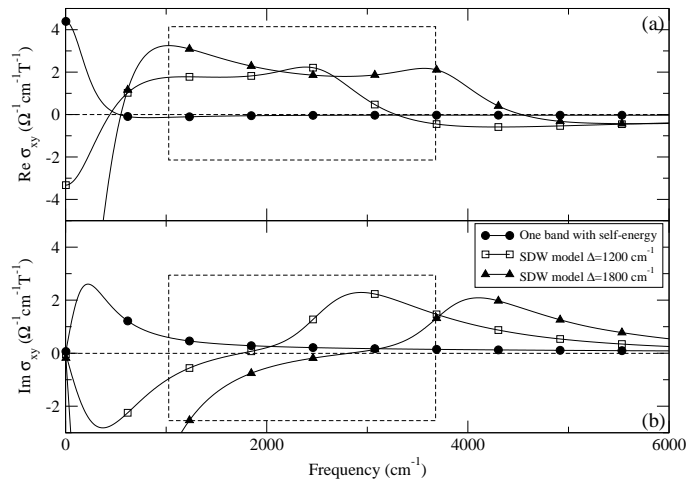}
\end{center}\caption{(a) Real and (b) imaginary part of $\sigma_{xy}$
as a function of frequency based on a spin density wave model
(SDW) with three different gap values. Note that the zero-crossing
diminishes with smaller input gap values $\Delta$ in the SDW model
as observed experimentally in figures \ref{Fig1} and \ref{Fig2}
(see dash boxes for better comparison).} \label{Fig3}
\end{figure}

To go beyond the Drude model and understand the entire set of
data, we now develop a spin density wave (SDW) model. We consider
a mean field theory of electrons moving on a square lattice with
dispersion
\begin{eqnarray}
\varepsilon_p= -2t(\cos p_x+\cos p_y)+\nonumber\\
4t^{\prime}\cos p_x \cos p_y -2t^{\prime\prime}(\cos 2p_x+\cos
2p_y)
\label{Eq1}
\end{eqnarray}
and parameters $t=0.38$eV=3056~cm$^{-1}$, $t^{\prime}=0.32t$ and
$t^{\prime\prime}=0.5t^{\prime}$, chosen to reproduce the Fermi
surface measured in the photoemission experiments, \cite{Armitage,
Millis}. The electrons are scattered by a commensurate $(\pi,\pi)$
spin density wave with scattering amplitude $\Delta$. The electric
and magnetic fields are represented by vector potentials and
coupled via the Peierls phase approximation, ${\vec p}\rightarrow
{\vec p}-\frac{e}{c}{\vec A}$. At the relevant carrier
concentration, the Fermi surface is such that the scattering
vector $(\pi,\pi)$ connects Fermi surface points, causing
reconstruction of the Fermi surface. At a doping of $x=0.15$, for
$\Delta$ less than $\Delta_c\approx 0.26$ eV, the calculated Fermi
surface exhibits both an electron-like pocket centered at
$(\pi,0)$ and hole-like pockets centered at $(\pm\pi/2,\pi/2)$ as
observed in photoemission \cite{Armitage}; for larger $\Delta$
only the electron pocket remains. The longitudinal and Hall
conductivities are calculated by direct evaluation of the Kubo
formula; for $\sigma_{xy}$ we used the expressions given by
Voruganti {\it et. al.} \cite{Voruganti92}. While this approach
does not capture the physics of the Mott transition, it does give
a reasonable picture of the effect of the SDW gap on the
quasiparticle properties.

Figure \ref{Fig3} shows the calculated IR Hall response for
$\Delta=0,$ $1200cm^{-1},$ $1800cm^{-1}$. For $\Delta=1200cm^{-1}$
and $1800cm^{-1}$ we used a constant scattering rate
$\gamma=0.2t=608cm^{-1}$.  The qualitative resemblance of the
calculated curves  to the data obtained for $x=0.12$ and $0.15$ is
striking (compared to figure \ref{Fig1}).  For non-zero $\Delta$ a
sign change occurs at a $\Delta$-dependent frequency and
correspondingly, in the range of frequencies actually measured,
Re($\sigma_{xy}$) becomes larger as $\Delta$ increases and two
sign changes occur; one at high and one at low frequency. The
zero-crossing seen in the Im($\sigma_{xy}$) may be understood as
follows: the SDW model predicts a Fermi surface made up of
hole-like and electron-like pockets. Im$(\sigma_{xy})$ is negative
(electron-like) at low frequencies where the large electron-like
pocket dominates the transport. However, at high frequency
(frequencies much higher than the SDW gap $\Delta$) the response
must revert to the hole-like response of the underlying hole-like
band structure. Similarly, the changes observed  with increasing
temperature are clearly similar to those occurring as the gap is
reduced in the model calculation. Although the qualitative
correspondence between calculation and data is striking an
important difference exists: the magnitude of the calculated Hall
response is bigger than the measured values by factors of the
order of 5. We believe this difference is the signature of the
suppression of AC charge response by Mott physics. Understanding
this in more detail is an important open problem.

Figure \ref{Fig3} also shows a calculation of the IR Hall response
for $\Delta=0$; this is to be compared to the data obtained for
$x=0.18$. For this doping we have used a frequency dependent
scattering model previously used to fit the $x=0.18$ $\sigma_{xx}$
data \cite{Voruganti92}. The rough magnitude and frequency
dependence are in qualitative accord with the measurement, but the
incorrect position of the zero-crossing in Re($\sigma_{xy}$)
demonstrates that despite its success in fitting the longitudinal
conductivity the extended Drude model is an inadequate description
of transport even in overdoped PCCO.

It is useful to consider the small measured values of
$\sigma_{xy}$ in terms the general issue of optical sum rules. For
the longitudinal conductivity $\sigma_{xx}$ the appropriate
partial sum is
$K(\Omega)=\int_{0}^{\Omega}\frac{2}{\pi}$Re$\sigma_{xx}(\omega)d\omega$.
In high temperature superconductors, theoretical and experimental
evidence suggests that for 0.1eV $\lesssim \Omega \ll $ 1eV , i.e.
above the Drude peak but well below the interband and upper
Hubbard band features, the partial integral is proportional to the
product of the doped hole density per Cu atom and the band theory
conduction band kinetic energy $K_{band}\approx$ 0.4eV : for
0.18-doped PCCO, $K$(0.4eV ) = $K^{doped-hole} \approx x K_{band}$
\cite{Millis}. For frequencies below this value the extended Drude
parametrization
$\sigma_{xx}(\omega)=K^{doped-hole}/(\gamma_{xx}(\omega)-i\omega(1+\lambda_{xx}(\omega)))$
is physically meaningful. For the Hall conductivity, the
appropriate partial sum is
$K(\Omega)=\int_{0}^{\Omega}\frac{2}{\pi}\omega $Im$
\sigma_{xy}(\omega)d\omega$ \cite{Drew} and the corresponding
extended Drude parametrization is
$\sigma_{xy}(\omega)=S^{doped-hole}/(\gamma_{xy}(\omega)-i\omega(1+\lambda_{xy}(\omega)))^2$.
For the low doped samples the effects of the SDW gap extends
through our measurement range.  Therefore we focus on the
overdoped x=0.18 sample for which there are no obvious high
frequency SDW effects. We can estimate S(0.4eV, 0.1eV) =
$\int_{0.1eV}^{0.4eV}$ directly from the data, finding about
0.045$S_{Band}$. We may estimate the contribution arising from
below our measurement range in two independent ways; the two
methods yield consistent answers. First, we observe that the very
low frequency longitudinal conductivity is characterized by a
"Drude peak" of width $\gamma^*\approx 0.01eV $ and a Hall
resistance not too far from the band value, suggesting that at low
frequencies a picture of weakly scattered quasiparticles is
appropriate, so that $\gamma^*_{xy}=\gamma^*$, allowing us to
estimate the integral over the low frequency region as
$\int_{0}^{0.1eV}\frac{2}{\pi}\omega $Im$
\sigma_{xy}(\omega)d\omega \approx
S^{doped-hole}/(1+\lambda_{xy}(\omega=0))^2 \approx
\sigma_{xy}(\omega=0)/(\gamma^*)^2\approx 0.035 S_{band}$, leading
to the estimate S(0.4eV ) $\approx 0.08 S_{band}$. Second, model
calculations \cite{Lin} show that in the frequency range
$\Omega\sim$0.2eV the quantity
$(\Omega/$Im$(\sigma_{xy}(\Omega))^2$ is about a factor of two
larger than $S(\Omega)$. Using this factor of two and our data
gives the estimate S(0.4eV )$\approx0.1S_{Band}$. Thus, we
conclude that the IR hall data at x = 0.18 imply a suppression of
the Hall effect which is approximately a factor of two greater
than the suppression of the longitudinal conductivity, rasing the
possibility of different Mott renormalizations of the Hall and
longitudinal conductivities. We suggest that refining the
experimental estimates and extending them to other materials, as
well as developing an appropriate body of theoretical results, are
urgent open problems.

We have reported the transverse optical conductivity
$\sigma_{xy}(\omega)$ of the electron-doped cuprate
Pr$_{2-x}$Ce$_{x}$CuO$_{4}$. At low doping and temperatures, the
results are consistent with a spin density wave scenario which was
previously used to explain photoemission and standard optical
conductivity data. At high doping, we find no sign of this SDW
gap, however a spectral weight analysis shows that the material
could still manifest effects from this gap in its optical response
function.

The authors thank Dr. S. Dhar for RBS / Channeling measurements and
Y. Dagan for his initial work on film preparation using LSGO
substrates. We wish to acknowledge useful discussions with A. V.
Chubukov and H. Kontani. The work at the University of Maryland was
supported by NSF grants DMR-0352735 and DMR-0303112. The work at the
University at Buffalo was supported by the Research Corporation
Cottrell Scholar Award and NSF-CAREER-DMR0449899. A. J. Millis and
J. Lin acknowledge support from NSF DMR 0403167.

\end{document}